# NICMOS transmission spectroscopy of HD 189733b: controversy becomes confirmation

P. Deroo[1], M. R. Swain[1] and G. Vasisht[1]

[1] Jet Propulsion Laboratory, California Institute of Technology, 4800 Oak Drive Pasadena, CA 91103

ABSTRACT

Spectral features corresponding to methane and water opacity were reported based on spectroscopic observations of HD 189733b with Hubble/NICMOS. Recently, these data, and other NICMOS exoplanet spectroscopy measurements, have been reexamined in Gibson et al. 2010, who claim that the features in the transmission spectra are due to uncorrected systematic errors and not molecular opacities. We examine the methods used by the Gibson team and show that, contrary to their claim, their results for the transmission spectrum of HD 189733b are in fact in agreement with the original results. In the case of HD 189733b, the most significant problem with the Gibson approach is a poorly determined instrument model, which causes (1) an increase in the formal uncertainty and (2) instability in the minimization process; although Gibson et al. do recover the correct spectrum, they cannot identify it due to the problems caused by a poorly determined instrument model. In the case of XO-1b, the Gibson method is fundamentally flawed because they omit the most important parameters from the instrument model. For HD 189733b, the Gibson team did not omit these parameters, which explains why they are able to reproduce previous results in this case, although with poor SNR.

## 1 INTRODUCTION

Measurements with the NICMOS instrument on the Hubble space telescope demonstrated that molecular spectroscopy of exoplanet atmospheres was possible and provided the first detection of methane in a planet orbiting another star (Swain, Tinetti & Vasisht 2008). From this initial result, molecular spectroscopy of exoplanet atmospheres has grown rapidly using Spitzer (Grillmair et al. 2008), Hubble (Swain et al. 2009a; Swain et al. 2009b), and ground-based measurements (Swain et al. 2010; Snellen et al. 2010; Thatte et al. 2010). The alkali element sodium has also been detected in exoplanet atmospheres using Hubble, (Charbonneau et al. 2002) and ground-based measurements confirmed this detection (Redfield et al. 2008; Snellen et al. 2008). Spectroscopic measurements reporting an absence of atomic or molecular features have also been reported (Pont et al. 2008; Pont et al. 2009), with useful constraints being placed by the measurements. At the present time, numerous teams have exoplanet spectroscopy programs in place, and at least one purpose-built instrument is being constructed. In short, the era of exoplanet characterization via spectroscopy is at hand.

Although the exoplanet spectroscopy field is broadening rapidly, the role of measurements with Hubble/NICMOS is currently unique. Operating in the near-IR, HST/NICMOS has produced published observations on four exoplanet systems and molecules were detected in three; transmission and emission spectra were obtained, and the measurement show that water, methane, and carbon dioxide are routinely present in hot-Jupiter atmospheres. Simply put, HST/NICMOS has made an enormous contribution to the area of exoplanet spectroscopy; the fact that the NICMOS instrument is currently not available, with no specific plans to return it to operation, is certainly a setback for the exoplanet community. Given the large and unique impact of the NICMOS instrument, independent confirmation of results is a high priority.

In the recent paper by Gibson, Pont & Aigrain (2010; hereafter GPA10), the authors reanalyzed transmission spectroscopic measurements and advance the hypothesis that residual systematic errors, not molecular opacities, are responsible for the spectral modulation in all of the published NICMOS exoplanet spectra. They investigate this hypothesis with a reanalysis of NICMOS transmission spectra for HD 189733b (Swain, Vasisht & Tinetti 2008; hereafter SVT08), GJ 436b (Pont et al. 2009), and XO-1b (Tinetti et al. 2010). Based on this reanalysis, GPA10 claim that (1) the previously published results significantly underestimate the measurement uncertainty and that (2) different decorrelation methods produce a different result. By deduction, they then conclude that the previously reported spectral modulation is explainable by residual systematic errors. As our team has previously published the transmission spectra of HD 189733b and XO-1b, we examined the GPA10 results for these two data sets in detail and explored the significant differences in their analysis that lead to their controversial conclusion.

## 2 OBSERVATIONS & METHODS

NICMOS observations of HD 189733b and XO-1b have been previously described in SVT08, Tinetti et al. 2010 and GPA10. For completeness, we summarize the NICMOS observations and calibration methods here.

The observational scheme for NICMOS exoplanet spectroscopy consists of a series of spectrophotometric observations, $F(t)$, centered on the exoplanet eclipse. For a transmission spectrum, a transit of the planet is observed, and by measuring the difference between in and out-of-eclipse (through modeling with a theoretical light-curve), the exoplanet spectrum is measured. For HD 189733b, two Hubble orbits prior to the transit and two orbits after the transit are observed to establish the out-of-eclipse baseline. For XO-1b, two orbits before and one after transit are observed for the same reason.

The Hubble/NICMOS instrument, like any instrument, introduces systematic effects that need to be carefully corrected to obtain the best possible spectro-photometry. This is done by establishing a model for the instrument systematics (instrument model) through a combination of seven instrument state vectors, collectively identified as $\Psi_i$ where the index *i* identifies the individual state vector. The calibration approach has been described in detail in SVT08 and updated in Swain et al. 2009a and 2009b; GPA10 implement a very similar method. The instrument state vectors are x position, *x*; y position, *y*; the angle of the spectrum on the detector, $\theta$; the full width half maximum of the point spread function, *fwhm*; the Hubble telescope's orbital phase, $\phi$; orbital phase squared, $\phi^2$; and temperature, *T*. With these state vectors, the observed spectrophometry is modeled using (1) an instrument model constructed using a linear combination of the state vectors and (2) a light curve model with the radius ratio between planet and star as a free parameter. In SVT08 and GPA10, the two components of the spectrophotometric model are determined separately by establishing the instrument model using the out-of-eclipse data and then applying the interpolated model to the in-eclipse portion of the light curve. For XO-1b in Tinetti et al. 2010, both components are determined simultaneously through a joint minimization over the entire time-domain (including in-eclipse). The minimization process for both cases minimizes the residuals $\varepsilon$ obtained by removing the eclipse light curve and the instrument model from the data:

$$\varepsilon(t) = F(t) - LM(d) \cdot \sum_i \beta_i \cdot \Psi_i(t) \qquad (1)$$

where the $\beta_I$ are the model coefficients for the instrument state vectors and *d* is the eclipse depth parameter for an idealized light curve model *LM,* which includes limb darkening. The minimization is done using either the Gauss-Markov method (SVT08) or using a downhill simplex method (for joint minimization; Tinetti et al. 2010), with both methods producing statistically identical results. This approach delivers an exoplanet spectrum by estimating the eclipse depth at every wavelength, and the error bars are determined by the fit residuals. GPA10 use a very similar approach to SVT08 and updates. Given the similar approach, the most instructive differentiation between both analyses is made by comparing their respective instrument models and the stability of the minimization (aka decorrelation), and it is here that significant differences exist.

## 4 DISCUSSION

The GPA10 paper contains several inaccuracies: some of these are relatively minor, while others are more serious. We focus here on explaining the crucial mistakes in GPA10 that led to their controversial claim that no molecular features have been detected in any NICMOS spectroscopy.

## 4.1 HD 189733b

The most striking aspect of the GPA10 result is that when they include all available data on the HD 189733b transit and implement a linear decorrelation (e.g. Eq. 1), they find an exoplanet transmission spectrum consistent with the SVT08 result (see Fig. 9 GPA10; reproduced here in Figure 1). *However, the GPA10 spectrum has much larger formal uncertainties – where do these uncertainties originate?* A detailed comparison shows that the GPA10 instrument model is poorly determined and is a significant source of error. A direct comparison of the SVT08 and the GPA10 instrument models is shown in Fig. 2 using published figures in both the SVT08 and GPA10 papers (a combination of Figure 7 in GPA10 and Figure 7 of the Supplementary Information of SVT08). Fig. 2 shows the photometric time series and instrument model for both publications; the photometry displays a similar scatter, but the GPA10 instrument model displays a scatter which is >3 times larger than SVT08. This cannot be the result of worse photon noise in GPA10. Based on the number of electrons and taking into account the analog-digital conversion gain of 6.5 (electrons/DN), the photon noise of the GPA10 photometry should be ~20 % better. Since GPA10 do not indicate what wavelength they used for their photometry example, we show in Fig. 3 the SVT08 spectrophotometry and instrument model for all wavelength channels; this clearly establishes that the SVT08 instrument model is at least 3 times less noisy than the example of GPA10. The consequence of a noisy instrument is significant, and the residuals obtained by removing the instrument model from the data will be significantly larger for GPA10. *The GPA10 finding of a similar transmission spectrum, with larger error bars, for HD 189733b is not the controversy the authors conclude; rather it is a confirmation of the SVT08 analysis by reanalyzing the same data, but with a poorly determined instrument model*. It is unfortunate that GPA10's failure to recognize the impact of a poorly determined instrument model led them to focus on dismissing NICMOS spectroscopy rather than acknowledging their fundamental agreement with previous results.

We now turn to consider the origin of the poorly determined instrument model used by GPA10. Since, the instrument model is constructed through a combination of the instrument state vectors, the noisier instrument model is likely the result of noisier instrument state vectors. We compared the instrument state vectors in GPA10 and SVT08, and found the most significant difference is in the determination of the rotation angle of the spectrum on the detector. GPA10's estimate for the angle is extremely noisy and is a factor ~ 5 noisier than that of SVT08. We compared the angle determination of GPA10 and SVT08 in Figure 4, and the scatter in the GPA10 estimate of the rotation angle is clearly identifiable as ~5 times greater than in the case of SVT08. In the case of GPA10, the significantly worse noise properties of the angle state vector directly propagate into the instrument model, and from there into the minimization process, ultimately affecting the measurement of the transmission spectrum. Interestingly, the rotation point for the spectrum lies ~ 2.1 μm (i.e. around the direct image position) and as a consequence, the angle state vector is less important for describing the instrument behavior for wavelengths in this region. The clear influence of the noise on the angle determination can be seen in the size of the error bars in the GPA10 exoplanet transmission spectrum (Figure 1), which follow a systematic pattern with small errors found around the rotation point and much larger error bars as distance from the rotation point increases (where the angle plays a prominent role in the decorrelation process). Thus, there is a self-consistent explanation for the large uncertainties reported by GPA10, and their results are due to a poorly determined estimate for the angle of the spectrum on the detector. Simply put, a noisy instrument model produces noisy results.

In their paper, GPA10 also explore the stability of the decorrelation process by removing portions of the data and by adding the optical state vectors squared to the decorrelation process. Decorrelation with portions of the data removed is a standard test for stability of the result, and extensive tests of this kind were done as part of the original SVT08 analysis and in the later publications on NICMOS spectroscopy (Swain et al. 2009a, Swain et al. 2009b, Tinetti et al. 2010). *In the case of SVT08, the exoplanet transmission spectrum was robust, so why did GPA10 find just the opposite?* The answer is that the noise introduced in the residuals by the poorly determined GPA10 instrument model requires all the data to marginally detect the exoplanet spectrum. Without all the data, the minimization process, which minimizes the $\chi^2$ difference between model and photometry, is too noisy to easily, and in a robust fashion, find the correct minimum. In addition, the noise in the instrument model may be systematic, which would make a $\chi^2$ minimization problematic when, like for GPA10, the noise in the instrument model becomes approximately equal to the noise in the spectrophotometry (see Figure 2). The version of the HD 189733b transmission spectrum GPA10 show in their Figure 28 is the result of one such "data removal" test that they perform. Given that:

1. the GPA10 data removal tests give inconsistent results (unlike SVT08),
2. the poorly determined instrument model used by GPA10 is influencing the result,
3. the spectrum in GPA10 Figure 28 is obtained by omitting one-third of the data, and
4. that GPA10 reproduce the original spectrum when they include all the usable data (albeit with larger errors due to a poorly determined instrument model), we conclude that the NICMOS spectrum shown in GPA10 Figure 28 contains no astrophysical information.

In the case of the non-linear decorrelation method (by which the GPA10 authors mean linear decorrelation with the inclusion of optical state vectors squared), GPA10 adds the same 5 instrument state vectors (including the noisy angle), but now squared to the minimization, and simultaneously increases the number of free parameters from 8 to 13 by "adding the terms $\Delta X^2$, $\Delta Y^2$, $W^2$, $\theta^2$ and $T^2$ to the state matrix". It is common knowledge that fitting with squared terms amplifies the noise, and this, combined with a significant increase in the number of free parameters, will make the minimization significantly less stable. Contrary to the claim in GPA10 that "there is no physical reason to believe the baseline flux should be modeled as a linear function", we note there is a clear and well established reason; at least 10 pixels are strongly irradiated in the spatial direction, and the shift of the spectrum on the detector is always less than 0.1 pixels for every wavelength. Thus the detector illumination pattern therefore changes by less than 1%, at which point almost any reasonable instrument response function can be approximated using a linear approach.

The poorly determined instrument model of GPA10 is the primary origin of the difference between both analyses of the HD 189733b data. This poorly determined instrument model influences the uncertainty of the exoplanet spectrum since the spectrophotometry is corrected by dividing by the instrument model. For GPA10, correcting the instrument systematics with a poorly determined instrument model increases the uncertainty in comparison to the SVT08 result. It is important to keep in mind that, when done properly, the signal-to-noise (SNR) per spectral channel of the molecular features is only about ~3.5 σ, so the factor ~3 increase in the uncertainty of the instrument model is crucial. With higher intrinsic SNR for the exoplanet spectral modulation, GPA10 could probably have gotten away with using a poorly determined instrument model. As it was, the combination of the poorly determined instrument model interacted with missing data and/or the non-linear decorrelation to produce inconsistent results. This illustrates the critical importance of well-determined instrument state vectors that permit, as in the case of SVT08, a stable result for the exoplanet spectrum.

4.2 XO-1b

Unlike the case of HD 189733b, where some modest level of due diligence is needed to determine the origin of the difference between the SVT08 and the GPA10 result, the case of XO-1b is straightforward. In GPA10, the authors declare they exclude three instrument state vectors ($\Delta X$, $\Delta Y$ and $\theta$) when constructing the instrument model. This is a crucial mistake; these instrument state vectors are not optional. Spectrophotometry with NICMOS is limited by inter-pixel and intra-pixel responsivity differences. The excluded instrument state vectors are the most important instrument state vectors for modeling the instrument behavior because they define the most important changes in the pixel-illumination. The GPA10 decision to exclude those state vectors is even more puzzling considering the authors end their XO-1 discussion by stating that $\theta$ is the most important instrument state vector and refer to the HD 189733b case as proof. The rational for exclusion of these state vectors, provided by GPA10, is that the spectrum moved by < 0.4 pixels. As with HD189733, the detector illumination changes for the XO-1 data are small, less than 4%, and for wavelengths close to the rotation point, the illumination functions changes by only 0.04 %. Given the continuity of the data, modeling the small changes as linear is valid and including higher order terms will amplify the noise. Since GPA10 do not include the most important state vectors for determining the instrument model, their disagreement with previous results is both understandable and irrelevant. The only conclusion that can be drawn from the GPA10 result for the XO-1b spectrum is that an incorrect instrument model will generate an incorrect result.

## 5 CONCLUSIONS

We have examined the GPA10 method in detail and have found the reasons for the disagreement between the GPA10 results and previous results by SVT08 and Tinetti et al. 2010. We show the claim made by GPA10, that the spectral modulation in Hubble/NICMOS spectra is due to residual instrument systematics, is unjustified due to significant errors in their method.

In the case of HD 189733b, a noisy and poorly determined instrument state vector, the angle of the spectrum on the detector, results in a noisy instrument model that, in turn, produces the large uncertainties in the resulting exoplanet spectrum. However, we note the HD 189733b spectrum obtained by GPA10, while noisy, is in fact consistent with the spectrum reported by SVT08. Unlike the SVT08 analysis, which found a stable result when portions of the data were removed, the poorly determined instrument model of GPA10 produces instability in the results when portions of the data set are removed or when a non-linear decorrelation is used. A poorly determined model for the instrument systematics fundamentally compromises the GPA10 results for HD 189733b. The case of XO-1b is different; here, the omission in GPA10 of three important instrument state vectors from the instrument model results in the (self-imposed) inability to correct the spectrophotometry and determine the exoplanet spectrum. In the case of HD 189733b, it is unfortunate that the failure by the GPA10 team to recognize their poor instrument model led the authors to ignore the fundamental agreement of their analysis with that of SVT08. Although clearly not the intent in GPA10, this agreement constitutes a confirmation of the original SVT08 result.

## Acknowledgements

The research was carried out at the Jet Propulsion Laboratory, California Institute of Technology, under a contract with the National Aeronautics and Space Administration.

## References

1. Charbonneau D., Brown T.M., Noyes R.W., Gilliland R.L., 2002, ApJ, 568, 377
2. Gibson N. P., Pont F., Aigrain S., 2010, preprint (astro-ph/ arXiv:1010.1753)
3. Grillmair C. J. et al., 2008, Nature 456, 767
4. Pont F., Knutson H., Gilliland R.L., Moutou C., Charbonneau D., 2008, MNRAS, 385, 109
5. Pont F., Gilliland R.L., Knutson H., Holman M., Charbonneau D., 2009, MNRAS, 393, 6
6. Redfield S., Endl M., Cochran W.D. Koesterke L., 2008, ApJ, 673, L87

7. Swain M.R., Vasisht G., Tinetti G., 2008, Nature 452, 329
8. Swain M.R., Vasisht G., Tinetti G., Bouwman J., Chen P., Yung Y., Deming D., Deroo P., 2009, ApJ, 690, L114
9. Swain M. R. et al., 2009, ApJ, 704, 1616
10. Swain M.R. et al. 2010, Nature, 463, 637
11. Snellen I.A.G, Albrecht S., de Mooi E.J.W., Le Poole R.S., 2008, A&A, 487, 357
12. Thatte A., Deroo P., Swain M.R., 2010, A&A (A&A preprint doi http://dx.doi.org/10.1051/0004-6361/201015148)
13. Tinetti G., Deroo P., Swain M.R., Griffith C.A., Vasisht G., Brown L.R., Burke C., McCullough P., 2010, ApJ, 712, L139

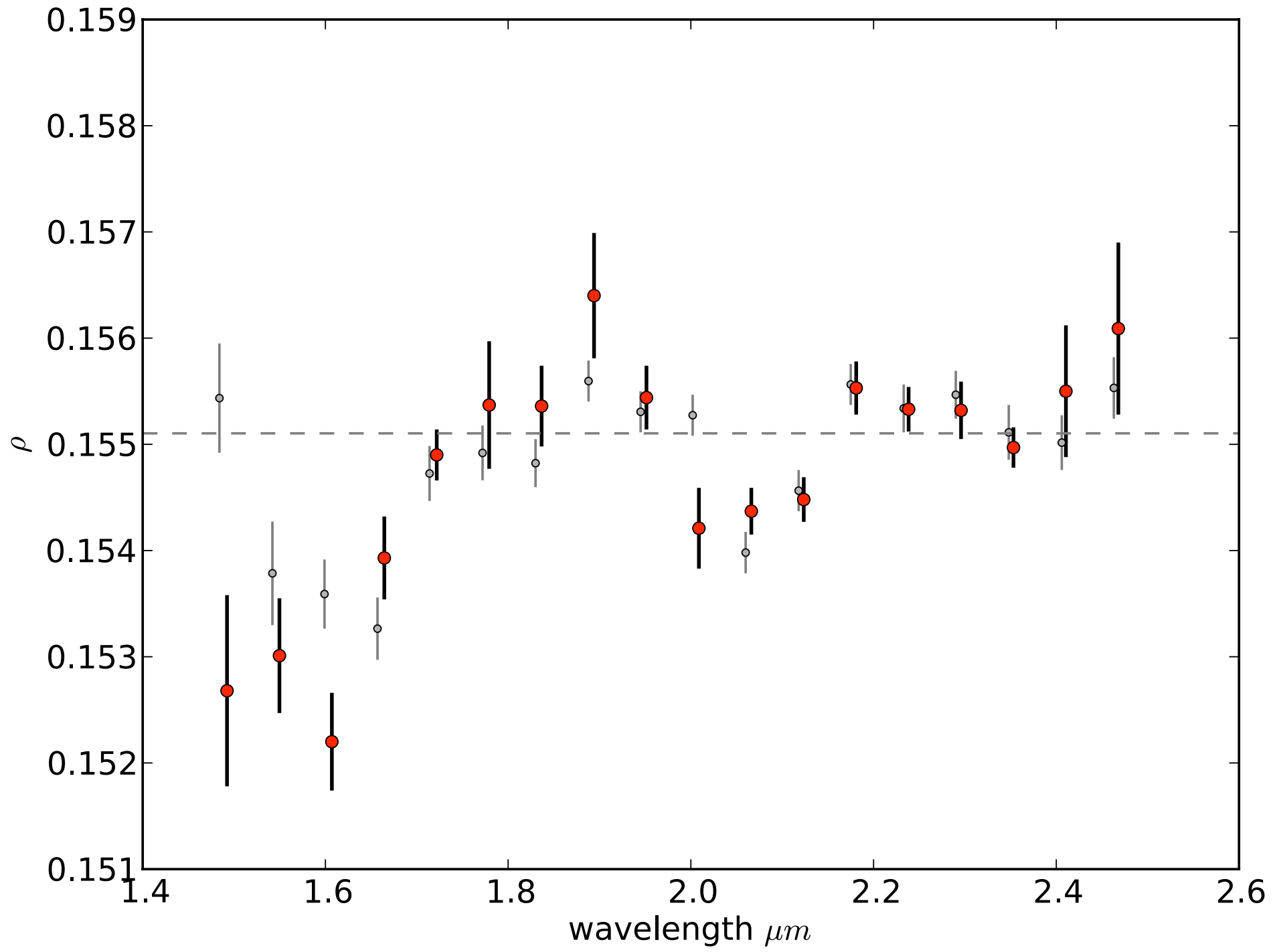

**Figure 1: Reproduction of Figure 9 in GPA10, which compares the transmission spectrum obtained in GPA10 (red) and SVT08 (grey) when using all available data and a linear decorrelation. Overall, the agreement between the results is good, but with larger uncertainties found by GPA10. Similar uncertainties are found around the rotation point of the spectrum (~ 2.1 μm), while the GPA10 error bars are larger elsewhere. In the test, we show that (1) the origin of the larger uncertainties in GPA10 lies in their significantly noisier instrument model, and (2) the noise in their instrument model is likely driven by a poor determination of the rotation angle of the spectrum. The rotation point for the spectrum is ~2.1 μm, and it is here, where the correction for rotation is smallest, that the GPA10 uncertainties are the smallest and are in good agreement with the SVT08 result.**

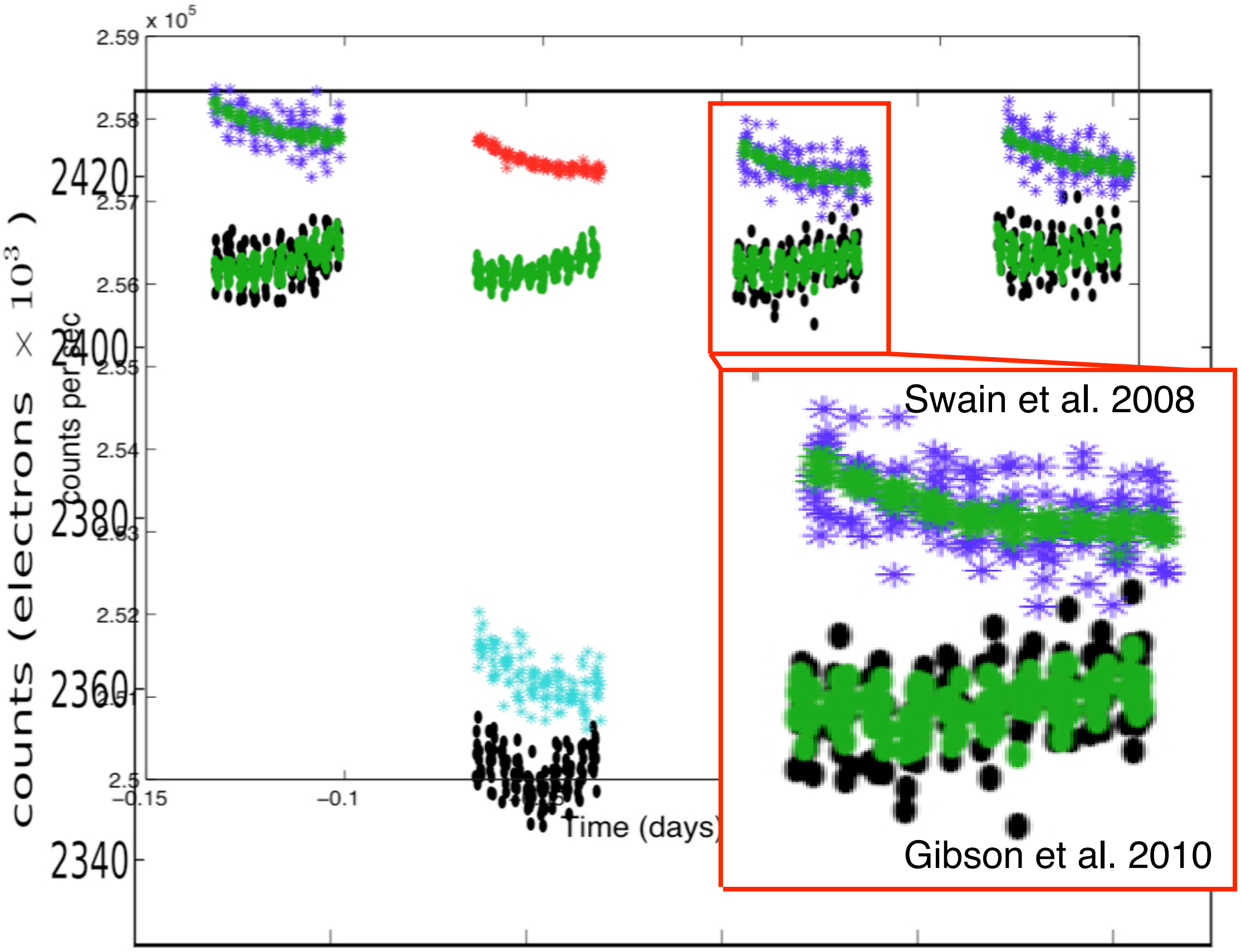

**Figure 2: Comparison of the instrument model in SVT08 and GPA10. This figure is the combination of Figure 7 in GPA10 and Figure 7 in the supplementary index of SVT08 over-plotted and rescaled to the same range (the scaling is such that the depth of the primary eclipse is the same). In the inset, we zoom to compare the instrument model of SVT08 and GPA10, which is shown in green in both publications. While the photometry in GPA10 and SVT08 both show similar scatter, the model for the instrument systematic effect in GPA10 is >3 times noisier than the model of SVT08. The noisy (poorly determined) instrument model of GPA10 is the origin of both the larger uncertainty and decorrelation instability reported by GPA10.**

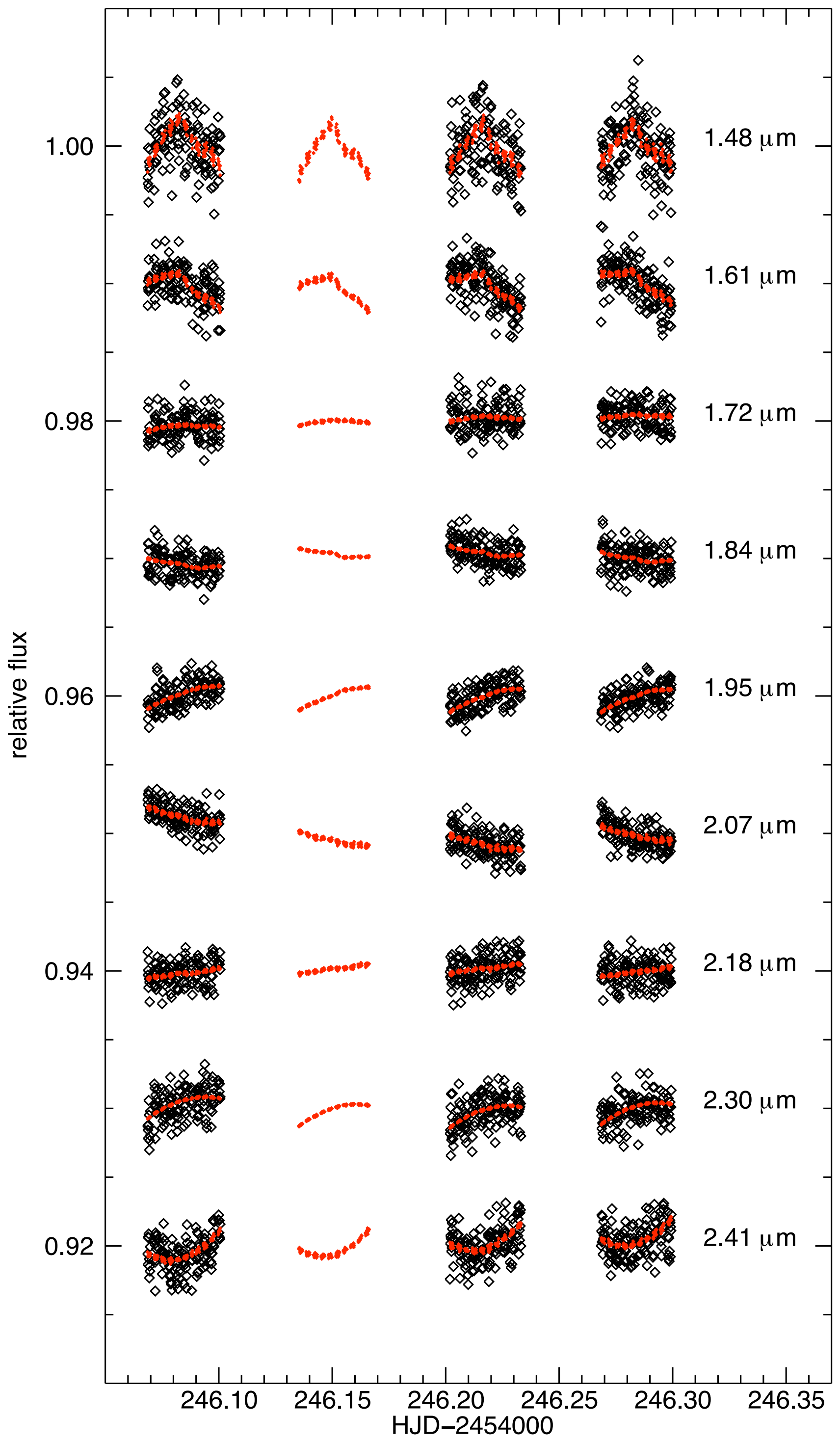

**Figure 3: The photometry (black) for individual spectral channels is compared to the instrument model (red) determined by SVT08. Clearly, none of the spectral channels has an instrument model as noisy as the GPA10 instrument model. In GPA10, the noise in the instrument model approximates the noise of the spectrophotometry (see Figure 1 this paper and Fig 7 GPA10).**

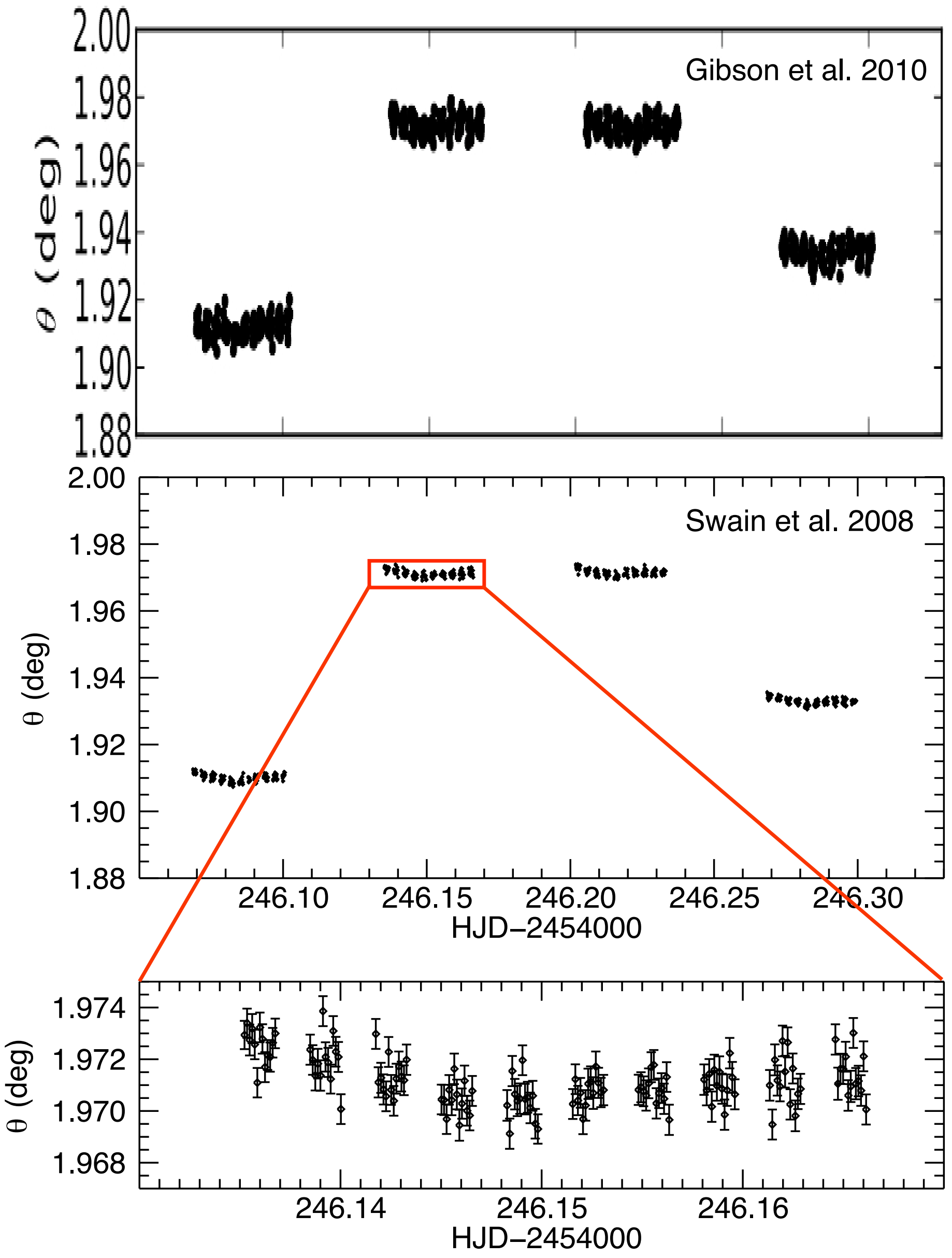

**Figure 4: The determination of rotation angle of the spectrum on the detector is compared between GPA10 (top panel) and SVT08 (middle and bottom panel). The top panel is a reproduction of Figure 6 in GPA10, while in the middle panel, the SVT08 values are shown using exactly the same scale for easy comparison. The angle determination in GPA10 displays about 5 times bigger scatter than in SVT08. In the bottom panel, we zoom in to part of the angle state vector determination such that the error bars become visible. The SVT08 1-σ error bar is ~$6 \cdot 10^{-4}$ deg, while the scatter in the angle determination of GPA10 spans about twice the full range of the bottom panel.**